\documentclass[12pt,english,dvips]{revtex4}
\usepackage{color}
\usepackage[T1]{fontenc}
\usepackage[ansinew]{inputenc}
\usepackage{amssymb}

\begin{document}

\begin{titlepage} \vspace{0.2in}

\begin{center} {\LARGE \bf

Geometrization of the electro-weak} \vspace*{0.3cm}

{\LARGE \bf model bosonic component}\\

\vspace*{1cm}
{\bf Francesco Cianfrani*, Giovanni Montani*}\\
\vspace*{1cm}
*ICRA---International Center for Relativistic Astrophysics\\
Dipartimento di Fisica (G9),\\
Universit\`a  di Roma, ``La Sapienza",\\
Piazzale Aldo Moro 5, 00185 Rome, Italy.\\
e-mail: montani@icra.it\\
        francesco.cianfrani@icra.it\\
\vspace*{1.8cm}

 \vspace*{1cm} \\

\vspace*{1cm}

{\bf   Abstract  \\ } \end{center} \indent In this work we develop a geometrical unification theory for gravity and the electro-weak model in a Kaluza-Klein approach; in particular, from the curvature dimensional reduction Einstein-Yang-Mills action is obtained. We consider two possible space-time manifolds: 1)$V^{4}\otimes S^{1}\otimes S^{2}$ where isospin doublets are identified with spinors;  2) $V^{4}\otimes S^{1}\otimes S^{3}$ in which both quarks and leptons doublets can be recast into the same spinor, such that the equal number of quark generations and leptonic families is explained. \\Finally a self-interacting complex scalar field is introduced to reproduce the spontaneous symmetry breaking mechanism; in this respect, at the end we get an Higgs fields whose two components have got opposite hypercharges.


\end{titlepage}

\section{Introduction}
The best results of modern physics are the two theories that describe the fundamental interactions, i.e. Standard Model and General Relativity. However, at the present no satisfying way to unify them is known, especially because of the mathematical and physical difficulties in the development of a Quantum General Relativity. Even if it is a general hope that the quantization of gravity would lead to the unification of all interactions, there is no indication that it will follow. For example Loop Quantum Gravity \cite{Ro}, one of the most promising candidate for a quantum General Relativity, does not contain other interactions, so it achieves no unification yet. Moreover, because cosmological sources we observe are only gravitational ones, on cosmological scale we can treat gauge carriers fields as perturbations. Therefore, the coexistence, in a unified picture, of a classical formulation for gravity and of a quantum theory for other interactions can be interpreted as a low energy effect.\\
The great achievement of General Relativity is the geometrization of the gravitational field; following this approach we can try to regard even others fields as geometric properties (geometrical unification models) and, in particular, as space-time metric components (Kaluza-Klein theories). This issue, obviously, implies extra geometrical degrees of freedom, that can be introduced by virtue of extra-dimensions. To make the multidimensional model consistent with a four-dimensional phenomenology, we require the unobservability for this additional coordinates and, due to quantum uncertainty, it can be reached by taking a compactified extra-space. It is also possible to consider non-compact Kaluza-Klein models, as in braneworld scenario \cite{RS99}, and they found applications in string theory; however, in those models a strange kind of unification arise, in fact while the gravitational field propagates in the bulk, the other interactions are restricted to the four-dimensional brane.\\
The oldest model with non-gravitational fields in the metric is the original Kaluza-Klein one \cite{1} \cite{2} \cite{3}, which deals with the unification of gravitational and electromagnetic interactions in a space-time manifold $V^{4}\otimes S^{1}$. To extend this procedure to a generic Yang-Mills theory, a geometrical implementation of the gauge group and of its algebra have to be performed. The latter is easily achieved by the introduction of an homogeneous extra-space (for the definition of homogeneous space see \cite{La99}) and by considering its Killing vectors algebra. 
Instead, the role of group transformations is played by extra-dimensional translations; however, unless the above mentioned unobservability is taken into account, the right gauge transformations on fields are not reproduced \cite{Io}. At the end, to geometrize in a Kaluza-Klein approach a gauge interaction, an extra-space, whose Killing vectors algebra is the same as the gauge group one, is required (for a review see \cite{MKKT} \cite{OW97}).\\
In our work we reproduce in this way all the features of the electro-weak model, an issue failed by other multidimensional theories.  In particular, we derive not only free gauge bosons Lagrangian from the dimensional splitting of the multidimensional curvature, but also we show how their interaction with spinor fields can be obtained by the splitting of free Dirac Lagrangian. Therefore, we are able to give a completely geometric interpretation to the boson component and we have to introduce just free spinors.\\
Furthermore, the left- and right-handed four-dimensional fields, separately, as the lightest modes of an extra-coordinates expansion arise, while any their linear combination acquires a mass term, having the order of the compactification scale. So the need of a distinct treatment for the two chirality eigenstates becomes a low energy effect.\\   
There are two space-times suitable for our approach:
\begin{itemize} 
\item a 7-dimensional manifold $V^{4}\otimes S^{1}\otimes S^{2}$,
\item a 8-dimensional manifold $V^{4}\otimes S^{1}\otimes S^{3}$. 
\end{itemize}
In the first case, we deal with eight-components spinors, so we assign a geometric meaning to the isospin doublet.\\  
In the second case, the number of dimensions of the gauge group is the same as that of the extra-space; from this condition we get the identification of gauge charges with extra-components of the fields momentum and the equality of the number of leptonic families and quark generations.\\ Moreover, as an extension respect to canonical Kaluza-Klein theories, a dependence by some four-dimensional scalar fields ($\alpha^{m}$) for the extra-space metric is considered.\\ At the end, the multidimensional analogous of the Higgs boson is introduced; in particular, we define it so that its two components have got opposite hypercharges. In this way it is possible to realize the spontaneous symmetry breaking to the $U(1)$ electro-magnetic group and to add, in the Lagrangian density, mass terms for all particles.\\
In particular, the work starts in section II with a short review of the electro-weak model, while in section III we define the space-time manifold and show how the Einstein-Hilbert action reduction leads to the Einstein-Yang-Mills one and to the terms describing $\alpha$ dynamics; in section IV, for matter fields we illustrate the form of the extra-coordinates dependence able to give the right gauge transformations and the conservation of gauge charges. In section V we apply the results of the previous analysis to the space-time manifolds $V^{4}\otimes S^{1}\otimes S^{2}$ and $V^{4}\otimes S^{1}\otimes S^{3}$, where we recast the observed four-dimensional fields into eight and sixteen components spinors, respectively. Section VI deals with the introduction of the scalar field responsible of the $SU(2)\otimes U(1)$ symmetry breaking, while in section VII brief concluding remarks follow.  

\section{Electro-weak model}
The electro-weak model is a $SU(2)\otimes U(1)$ gauge theory; $SU(2)$
transformations act only on left-handed components in the following way
\begin{equation}
\psi'_{L}=(I+\frac{i}{2}g\delta\omega^{i}T_{i})\psi_{L}
\end{equation}
where $\delta\omega^{i}$ are arbitrary infinitesimal functions, g is group's coupling constants, $T_{i}$ are group's generators (Pauli matrices) and $\psi_{L}$ is constituted by left-handed leptons (quarks) fields of the same family (generation)
\begin{equation}
\psi_{lL}=\left(\begin{array}{c} \nu_{lL} \\ l_{L} \end{array}\right)\qquad\psi^{}_{qL}=\left(\begin{array}{c} u_{qL} \\ d_{qL} \end{array}\right).
\end{equation}
The conserved charges associated to the $SU(2)$ invariance are the three components of the weak isospin
\begin{equation}
I_{i}=\int_{E^{3}}d^{3}x \psi^{\dag}_{l}T_{i}\psi_{l}.
\end{equation}
U(1) transformations act also on right-handed states, in particular the infinitesimal transformation law for matter fields is the following one
\begin{equation}
\psi'=(I+ig'y_{\psi}\delta\omega^{0}I)\psi
\end{equation}
where g' is the U(1) coupling constant and the hypercharge $y_{\psi}$ depends on the field and of the chirality state. In fact, we impose that the weak hypercharge, the conserved charge associated to these transformations
\begin{equation}
Y=y\int d^{3}x \psi^{\dag}\psi,
\end{equation}
is related to the electric charge and the third component of the weak isospin by the following relation
\begin{equation}
\frac{Q}{e}=Y+I_{3}\label{q}.
\end{equation}
By substituting ordinary derivatives with the ones containing gauge connections and by introducing gauge bosons free terms, the $SU(2)\otimes U(1)$ invariant Lagrangian density from the Dirac one is developed, i.e.  
\begin{eqnarray*}
\Lambda=\sum_{l,q=1}^{3}\frac{i\hbar c}{2}\bigg[D^{\dag}_{\mu}\bar{\psi}_{lL}\gamma^{\mu}\psi_{lL}-\bar{\psi}_{lL}\gamma^{\mu}D_{\mu}\psi_{lL}+D^{\dag}_{\mu}\bar{\nu}_{lR}\gamma^{\mu}\nu_{lR}-\bar{\nu}_{lR}\gamma^{\mu}D_{\mu}\nu_{lR}+
D^{\dag}_{\mu}\bar{l}_{R}\gamma^{\mu}l_{R}-\\-\bar{l}_{R}\gamma^{\mu}D_{\mu}l_{R}\bigg]+\sum_{g=1}^{3}\frac{i\hbar
c}{2}\bigg[D^{\dag}_{\mu}\bar{\psi}_{qL}\gamma^{\mu}\psi_{qL}-\bar{\psi}_{qL}\gamma^{\mu}D_{\mu}\psi_{qL}+D^{\dag}_{\mu}\bar{u}_{qR}\gamma^{\mu}u_{qR}-\\-\bar{u}_{qR}\gamma^{\mu}D_{\mu}u_{qR}+
D^{\dag}_{\mu}\bar{d}_{qR}\gamma^{\mu}d_{qR}-\bar{d}_{qR}\gamma^{\mu}D_{\mu}d_{qR}\bigg]
-\frac{1}{4}B_{\mu\nu}B^{\mu\nu}-\frac{1}{4}G_{i\mu\nu}G_{i}^{\mu\nu}
\end{eqnarray*}
where
\begin{eqnarray*}
B_{\mu\nu}=\partial_{\nu}B_{\mu}-\partial_{\mu}B_{\nu} \\
G^{i}_{\mu\nu}=\partial_{\nu}W^{i}_{\mu}-\partial_{\mu}W^{i}_{\nu}+gc^{i}_{jk}W^{j}_{\mu}W^{k}_{\nu}\\
D_{\mu}\psi_{lL}=[\partial_{\mu}+\frac{1}{2}ig\tau_{i}W^{i}_{\mu}-i\frac{g}{2}'IB_{\mu}]\psi_{lL}\\
D_{\mu}l_{R}=[\partial_{\mu}-ig'B_{\mu}]l_{R}\\
D_{\mu}\nu_{lR}=\partial_{\mu}\nu_{lR}\\
D_{\mu}\psi_{qL}=[\partial_{\mu}+\frac{i}{2}g\tau_{i}W^{i}_{\mu}+i\frac{g}{6}'IB_{\mu}]\psi_{qL}\\
D_{\mu}u_{qR}=[\partial_{\mu}+i\frac{2}{3}g'B_{\mu}]u_{qR}\\
D_{\mu}u_{qR}=[\partial_{\mu}-i\frac{1}{3}g'B_{\mu}]u_{qR}.\\
\end{eqnarray*}
Bosons we observe are related to $SU(2)\otimes U(1)$ gauge ones by the transformations 
\begin{equation}
\left\{\begin{array}{c}B_{\mu}=-\sin\theta_{w}
Z_{\mu}+\cos\theta_{w} A_{\mu}\\\\
W^{1}_{\mu}=\frac{1}{\sqrt{2}}(W^{+}+W^{-})\\\\
W^{2}_{\mu}=\frac{i}{\sqrt{2}}(W^{+}-W^{-})\\\\
W^{3}_{\mu}=\cos\theta_{w}Z_{\mu}+\sin\theta_{w}A_{\mu}
\end{array}\right..
\end{equation}
In order to obtain a spontaneous $SU(2)\otimes U(1)$ symmetry breaking a Higgs scalar field, which is an isospin doublet and has got $\frac{1}{2}$ hypercharge, is introduced with a suitable Lagrangian density, i.e.
\begin{eqnarray}
\Lambda_{\phi}=\frac{1}{2}[D_{\mu}\phi]^{\dag}[D^{\mu}\phi]-\mu^{2}\phi^{\dag}\phi-\lambda[\phi^{\dag}\phi]^{2}-\sum_{l}g_{l}[\bar{\psi}_{lL}l_{R}\phi+\phi^{\dag}\bar{l}_{R}\psi_{lL}]-g_{\nu_{l}}[\bar{\psi}_{lL}\nu_{lR}\widetilde{\phi}+\widetilde{\phi}^{\dag}\bar{\nu}_{lR}\psi_{lL}]-\nonumber\\-\sum_{q}g_{uq}[\bar{\psi}_{qL}u_{qR}\phi+\phi^{\dag}\bar{u}_{qR}\psi_{qL}]-g_{dq}[\bar{\psi}_{qL}d_{qR}\widetilde{\phi}+\widetilde{\phi}^{\dag}\bar{d}_{qR}\psi_{qL}]\label{mass}\qquad\qquad\qquad\qquad\qquad\qquad\qquad\qquad
\end{eqnarray}
with $\mu^{2}<0$, $\lambda >0$ and $\widetilde{\phi}=-i(\phi^{\dag}T_{2})^{T}$. The Higgs field realizes a spontaneous symmetry breaking, which deals to the conservation of only the electric charge, by fixing its expectation value in the vacuum as the following one
\begin{equation}
\phi_{0}=\left(\begin{array}{c} 0 \\ \frac{v}{\sqrt{2}}
\end{array}\right).
\end{equation}
In this way, a unified picture for weak and electro-magnetic interaction is achieved; furthermore, a mechanism by which massive particles arise is obtained, even if the field responsible of this process have not been detected yet.  

\section{Geometrization of the electro-weak bosonic component}
Let us consider a space-time manifold 
\begin{equation}
V^{n}=V^{4}\otimes B^{k}
\end{equation}
where $V^{4}$ is the ordinary 4-dimensional space-time, with coordinates $x^{\mu} (\mu=0,\ldots,3)$, and $B^{k}$ stands for the extra-dimensional one. The space $B^{k}$ is introduced in such a way that its Killing vectors reproduce the Lie algebra of a $SU(2)\otimes U(1)$ group, i.e.
\begin{equation}
\xi^{n}_{M}\partial_{n}
\xi_{M}^{m}-\xi^{n}_{M}\partial_{n}
\xi_{\bar{N}}^{m}=C^{P}_{NM}\xi^{m}_{P}
\end{equation}
where $C^{P}_{NM}$ are $SU(2)\otimes U(1)$ structure constants.\\
In particular, we have the following two possible choices for $B^{k}$ 
\begin{eqnarray}
i)S^{1}\otimes S^{2}\qquad\\
ii)S^{1}\otimes S^{3}\quad;
\end{eqnarray}
we will refer to coordinates on $S^{1}$ and $S^{2}$ or $S^{3}$ by $y^{0}$ and $y^{i} (i=1,..k-1)$, respectively, while $y^{m}\quad(m=0,..,k-1)$ will indicate both of them.\\
We develop a theory invariant under general four-dimensional coordinates transformations and under translations along $B^{k}$ 
\begin{equation}
\left\{\begin{array}{c} x'^{\mu}=x'^{\mu}(x^{\nu})\\
y'^{m}=y^{m}+\omega^{M}(x^{\nu})\xi^{m}_{M}(y^{n});
\end{array}\right.
\end{equation}
for the metric we take the following ansantz
\begin{equation}
j_{AB}=\left(\begin{array}{c|c}g_{\mu\nu}+\gamma_{mn}\xi^{m}_{M}\xi^{n}_{N}A^{M}_{\mu}A^{N}_{\nu}
& \gamma_{mn}\xi^{m}_{M}A^{M}_{\mu} \\\\
\hline\\
\gamma_{mn}\xi^{n}_{N}A^{N}_{\nu}
& \gamma_{mn}\end{array}\right)
\end{equation}
where $A^{M}_{\mu}$ are gauge bosons fields and $g_{\mu\nu}$ the four-dimensional metric \cite{Io}.\\
For the extra-dimensional metric, we assume
\begin{equation}
\gamma_{mn}(x;y)=\gamma_{mn}(x)\alpha^{m}(x)\alpha^{n}(x)
\end{equation}
(in the latter expression the indices m and n are not summed) with $\alpha=\alpha(x)$ some scalar fields that account for the dynamics of the extra-dimensional space. In order to forbid anisotropies of the space $S^{2}$ or $S^{3}$, we take $\alpha^{1}=\alpha^{2}(=\alpha^{3})=\alpha$, so allowing just size changes.\\
We note that in the 8-dimensional case the extra-space is 4-dimensional, so its dimensionality is the same as that of the gauge group; thus, Killing vectors can be chosen in such a way they are directly related to extra-dimensional n-bein vectors by the following relation
\begin{equation}
\xi^{(n)}_{n}=\alpha^{n}e^{(n)}_{n}
\end{equation}
where in the latter the index n is not summed.\\ 
We can carry on the dimensional reduction of the Einstein-Hilbert action in 4+k-dimensions as
\begin{equation}
S=-\frac{c^{3}}{16\pi G_{n}}\int_{V^{4}\otimes B^{k}} \sqrt{-j}{}^n\!R d^{4}xd^{k}y
\end{equation}  
and we get the geometrization of the gauge bosonic component, i.e. from the multidimensional curvature their Lagrangian density outcomes
\begin{eqnarray}
S=-\frac{c^{3}}{16\pi G}\int_{V^{4}}d^{4}x\sqrt{-g}
\bigg[R+R_{N}-2g^{\mu\nu}\sum_{n=0}^{k-1}\frac{\nabla_{\mu}\partial_{\nu}\alpha^{n}}{\alpha^{n}}-\nonumber\\-g^{\mu\nu}\sum_{n\neq
m=0}^{k-1}\frac{\partial_{\mu}\alpha^{m}}{\alpha^{m}}\frac{\partial_{\nu}\alpha^{n}}{\alpha^{n}}-\frac{1}{4}\sum_{M=1}^{3}(\alpha)^{2}G^{M}_{\mu\nu}G^{M\mu\nu}-\frac{1}{4}(\alpha')^{2}B^{M}_{\mu\nu}B^{M\mu\nu}
\bigg]\label{az}
\end{eqnarray}
where for Newton coupling constant and for the curvature term $R_{N}$ we have, respectively, (V is the volume of the extra-space)
\begin{eqnarray*}
G=\frac{G_{n}}{V}\qquad\qquad\quad\\
R_{N}=\frac{1}{4}\frac{1}{\alpha^{2}}C^{M}_{NT}C^{N}_{TM}.
\end{eqnarray*}
For details, we remand to our precedent work \cite{Io}.\\
Therefore in a Kaluza-Klein approach we carry on the geometrization by identifying gauge bosons with mist metric components. Unless the introduction of super-symmetries this procedure is not extensible to fermions, so to account for 
them we have to introduce spinor matter fields.

\section{Matter fields}
Any matter field is a not-geometric and, thus, an undesired term in a geometrical unification model; however, a remarkable result achieved in previous works  \cite{Io} \cite{M04} is the possibility to deal simply with free spinor fields, once we assume for them a suitable dependence on extra-coordinates.\\ 
In other words, we take the following matter fields 
\begin{equation}
\Psi(x;y)=\frac{1}{V}e^{-iT_{M}\lambda^{M}_{N}\Theta^{N}(y)}\psi(x)\label{mfield},
\end{equation}
where $T_{M}$ are gauge group's generators, $\Theta^{N}$ scalar densities of weight $\frac{1}{2}$ on the extra-space and $\psi$ the four-dimensional fields, while we define the constant matrix $\lambda$ by the following relation
\begin{equation} 
(\lambda^{-1})^{N}_{M}=\frac{1}{V}\int
\sqrt{-\gamma}\Big(\xi^{m}_{M}\partial_{m}\Theta^{N}\Big)d^{K}y.\label{lam}
\end{equation}
With such a dependence on extra-coordinates, the transformation of the field $\psi$ under translations along Killing vectors $\xi_{M}^{m}$, once the role of the unobservability is taken into account \cite{Io2}, is like the action of a gauge group, i.e. 
\begin{equation}
\psi'=\psi+i\omega^{M}T_{M}\psi.
\end{equation}
Moreover, with the same assumption we demonstrate the conservation of gauge charges; in particular, in the 8-dimensional case they can be regarded as extra-components of the fields momentum \cite{M04}.\\
The most remarkable result, that follows from the hypothesis (\ref{mfield}), is the geometrization of spinor gauge connections. In fact, starting from the free Dirac Lagrangian, the interaction with gauge bosons outcomes
\begin{equation}
\int_{B^{K}}\sqrt{-\gamma}d^{K}y
\bar{\Psi}\gamma^{(\mu)}D_{(\mu)}\Psi d^{k}y=\bar{\psi}\gamma^{(\mu)}e^{\mu}_{(\mu)}(D_{\mu}+iT_{M}A^{M}_{\mu})\psi.
\end{equation}
However, some additional terms, deriving from 
\begin{equation}
\label{mt}\int_{B^{K}}\sqrt{-\gamma}d^{K}y
\bar{\Psi}\gamma^{(m)}\partial_{(m)}\Psi d^{k}y,
\end{equation}
arise; they produce the standard Kaluza-Klein contribution to the spinor mass, of the same order as the compactification scale, and lead to very heavy four-dimensional particles. These particles cannot be described by a low energy theory.\\ 
Now, because the extra-dimensional $\gamma$ matrices are developed by the four-dimensional $\gamma_{5}$ one, we find that for the left-handed and right-handed fields the expression (\ref{mt}) vanishes.\\
The latter statement explains why, in a multidimensional theory, the four-dimension chirality eigenstate still have to be considered: they correspond to the lightest mode of an expansion in the extra-coordinates.\\
In conclusion, to account for fermions interacting with gauge bosons, we need just free left-handed and right handed spinors, where, again, we stress that the chirality we refer to is the four-dimensional one.  

\section{Spinors representation}
At this point, the introduction of fermions is based on the analysis of the previous section. In particular, the action of the $U(1)$ group is reproduced by adding a $y^{0}=\varphi$ dependent phase in front of the four dimensional spinors
\begin{equation}
\phi(x^{\mu})\rightarrow \widetilde{\Psi}=\frac{1}{\sqrt{L}}e^{in\varphi}\phi\qquad \varphi=[0;2\pi) 
\end{equation}
where $L$ stands for $S^{1}$ length and n is related to the hypercharge of the field $\phi$; in fact, under the infinitesimal translation $\varphi\rightarrow\varphi+f(x^{\mu})$, the transformation law for $\phi$ is 
\begin{equation}
\phi\rightarrow\phi+in f(x^{\mu})=\phi+i \frac{n}{6}g(x^{\mu}).
\end{equation}
From the latter expression, by imposing it reproduces the action of the hypercharge $U(1)$ group, the value of $n$ for each spinor is easily obtained 
\begin{equation}
n_{\psi}=6y_{\psi}.
\end{equation}
We observe that, in order to maintain the fundamental periodicity of $\widetilde{\Psi}(\varphi)$, n must be an integer; this gives a justification for the hypercharge spectrum observed.\\  
In an analogous but formally more complex way, for the $SU(2)$ group we make use of Pauli matrices and, according with the equation (\ref{mfield}), the form for the dependence on $S^{2}$ or $S^{3}$ coordinates is the following
\begin{equation}
\Psi=\frac{1}{\sqrt{V}}e^{-iT_{M}\lambda^{M}_{N}\Theta^{N}(y^{i})}\widetilde{\Psi}\quad M,N=1,2,3
\end{equation}
where $T_{M}$ are $SU(2)$ generators, $V$ stands for $S^{2}$ or $S^{3}$ volume and $\lambda$ obeys the relation (\ref{lam}).\\
Now, because the number of spinor components in a d-dimensional space-time is $2^{[\frac{d}{2}]}$ \cite{Av05}, we perform a distinct treatment for a 7-dimensional and a 8-dimensional space-time.\\
In the 7-dimensional case, we deal with eight components spinors, thus explaining the isospin doublet. In particular, we introduce the following four spinors for any leptonic family and quark generation
\begin{eqnarray}
\Psi_{lL}=\frac{1}{\sqrt{V}\sqrt{L}}e^{-i\sigma_{I}\lambda^{I}_{M}\Theta^{M}}
\left(\begin{array}{c}e^{in_{L}\varphi}\nu_{lL}\\
e^{in_{L}\varphi}l_{L}\end{array}\right)\quad
\Psi_{lR}=\frac{1}{\sqrt{V}\sqrt{L}}\left(\begin{array}{c}\nu_{lR}\\
e^{in_{lR}\varphi}l_{R}\end{array}\right)\quad l=1,2,3\\
\Psi_{qL}=\frac{1}{\sqrt{V}\sqrt{L}}e^{-i\sigma_{I}\lambda^{I}_{M}\Theta^{M}}
\left(\begin{array}{c}e^{in_{L}\varphi}\nu_{lL}\\
e^{in_{L}\varphi}l_{L}\end{array}\right)\quad
\Psi_{qR}=\frac{1}{\sqrt{V}\sqrt{L}}\left(\begin{array}{c}e^{in_{uR}\varphi}u_{lR}\\
e^{in_{dR}\varphi}d_{lR}\end{array}\right)\quad q=1,2,3
\label{spinors}
\end{eqnarray}
with obvious notation for four-dimensional fields. Thus, we include even the right handed fields of leptons (quarks) of the same family (generation) in the same spinor and this assumption is consistent with their different interaction properties, since we assume for them a different dependence on the coordinate $\varphi$.\\
In the 8-dimensional case spinors have got 16 components and this allow us to reproduce all Standard Model particles from the following six ones
\begin{eqnarray}
\Psi_{lL}=\frac{1}{\sqrt{V}\sqrt{L}}e^{-iT_{I}\lambda^{I}_{M}\Theta^{M}}
\left(\begin{array}{c}e^{in_{L}\varphi}\nu_{lL}\\
e^{in_{L}\varphi}l_{L}\\e^{in_{qL}\varphi}u_{lL}\\
e^{in_{qL}\varphi}d_{lL}\end{array}\right)\quad
\Psi_{lR}=\frac{1}{\sqrt{V}\sqrt{L}}\left(\begin{array}{c}\nu_{lR}\\
e^{in_{lR}\varphi}l_{R}\\e^{in_{uR}\varphi}u_{lR}\\
e^{in_{dR}\varphi}d_{lR}\end{array}\right)\quad l=1,2,3\label{spinors8}
\end{eqnarray}
being 
\begin{equation}
T_{I}=\left(\begin{array}{cc} \sigma_{I} & 0
\\ 0 & \sigma_{I} \end{array}\right)
\end{equation}
where we stress that Pauli matrices $\sigma_{I}$ act on isospin doublets.\\
Therefore, in a space-time $V^{4}\otimes S^{1}\otimes S^{3}$ we propose a model in which quarks and leptons are the components of the same spinor, but the Lorentz symmetries connecting them are broken by the compactification. The relic features of such a statement are \begin{itemize}\item i)common properties under the action of (what a fuor dimensional observer interpret as) the gauge group, \item the equality between the number of leptonic families and quark generations.\end{itemize}
In both 7- and 8-dimensional cases, from the dimensional splitting of Dirac Lagrangian for such spinors, we also get some relations between $\alpha$ and $\alpha'$ and electro-weak coupling constants $g$ and $g'$, so an estimate of extra-dimensions lengths, i.e.
\begin{equation}
\alpha^{2}=16\pi G\bigg(\frac{\hbar}{gc}\bigg)^{2}\qquad
\alpha'^{2}=16\pi G\bigg(\frac{\hbar}{g'c}\bigg)^{2}
\end{equation}
\begin{equation}
\alpha=0.18\times 10^{-31}cm\qquad
\alpha'=0.33\times10^{-31}cm.
\end{equation}
These equalities show that if $\alpha$ and $\alpha'$ are not constants, so if the extra-dimensional space changes with space and time (as we expect in a geometrodynamics theory), coupling constants acquire a dependence on (four-dimensional) space-time coordinates, like in Dirac hypothesis on large numbers \cite{Dir} or in the Brans-Dicke theory \cite{BD}.\\ 

\section{Spontaneous symmetry breaking}
Now, we have to reproduce the spontaneous symmetry breaking mechanism, to account for the only (electro-magnetic) symmetry we find in the physical world and to attribute masses to particles.\\
In a straightforward way we can just introduce a two components complex scalar field, i.e. 
\begin{equation}
\Phi=\left(\begin{array}{c} \Phi_{1}
\\ \Phi_{2}
\end{array}\right)=\frac{1}{\sqrt{V}}e^{-i\sigma_{I}\lambda^{I}_{M}\Theta^{M}}\left(\begin{array}{c}e^{-in_{\phi}\varphi}\phi_{1}
\\ e^{in_{\phi}\varphi}\phi_{2}
\end{array}\right)\quad n_{\phi}=3
\end{equation} 
and so the four-dimensional one associated 
\begin{equation}
\phi=\left(\begin{array}{c}\phi_{1} \\ \phi_{2}
\end{array}\right)
\end{equation}
has two weak isospin doublet components $\phi_{1}$ and $\phi_{2}$, with hypercharges $-\frac{1}{2}$ and $\frac{1}{2}$,  respectively.\\
We take a Lagrangian density with a Higgs potential
\begin{equation}
\Lambda_{\Phi}=\frac{1}{2}\eta^{(A)(B)}\partial_{(A)}\Phi^{\dag}\partial_{(B)}\Phi
-\mu^{2}\Phi^{\dag}\Phi-\lambda(\Phi^{\dag}\Phi)^{2} 
\end{equation}
and, after the dimensional splitting, we get the action for the Higgs boson
\begin{eqnarray}
S_{\Phi}=\frac{1}{c}\int_{V^{4}} d^{4}x \sqrt{-g}
\bigg[\frac{1}{2}g^{\mu\nu}(D_{\mu}\phi)^{\dag}D_{\nu}\phi-
(\mu^{2}-G)\phi^{\dag}\phi-\lambda(\phi^{\dag}\phi)^{2}\bigg]
\end{eqnarray}
unless an additional mass term $G\simeq\frac{1}{\alpha^{2}}\simeq 10^{19}GeV$, which would imply an extremely accurate fine tuning on the potential parameters $\mu$ and $\lambda$ to get a mass $m_{H}\sim 200GeV$.\\ 
Now, we have to fix a suitable vacuum expectation value to get the spontaneous symmetry breaking, but because for $\phi$ the hypercharge generator is opposite to the weak isospin third component one, from (\ref{q}) we have that any vacuum state is invariant under the electric charge symmetry.\\
Therefore, in the unitary gauge, we rewrite $\phi$ around the vacuum as
\begin{equation}
\phi=\left(\begin{array}{c} v_{1}+\sigma(x) \\ v_{2} \end{array}\right) \quad v_{1}=v\cos\theta\quad v_{2}=v\sin\theta.
\end{equation}
In order to obtain massive fermions, a distinct treatment between the 7-dimensional and the 8-dimensional scenario is required.\\
In the first case, the form of fermions mass terms in the Lagrangian is the following one 
\begin{eqnarray}
\Lambda_{\Psi\phi}=\sum_{l=1}^{3}\big[g_{l}(\bar{\Psi}_{lL}\Phi\Psi_{lR}+\bar{\Psi}_{lR}\Phi^{\dag}\Psi_{lL})\big]
+\sum_{q=1}^{3}\big[g_{u_{q}}(\bar{\Psi}_{qL}\Phi\Psi_{qR}+\bar{\Psi}_{qR}\Phi^{\dag}\Psi_{qL})\big]
\end{eqnarray}
where we define the invariant product of two spinors and the field $\Phi$ as 
\begin{equation}
\bar{\Psi}_{1}\Phi\Psi_{2}=\sum_{r=1}^{4}[(\bar{\Psi}_{1})_{r}\Phi_{1}(\Psi_{2})_{r}+(\bar{\Psi}_{1})_{r+4}\Phi_{2}(\Psi_{2})_{r+4}].
\end{equation}
In the second case, we deal with only one constant g, i.e.
\begin{equation}
\Lambda_{\Psi\phi}=g\sum_{l=1}^{3}[\bar{\Psi}_{lL}\Phi\Psi_{lR}+\bar{\Psi}_{lR}\Phi^{\dag}\Psi_{lL}]
\end{equation}
being
\begin{eqnarray*}
\bar{\Psi}_{1}\Phi\Psi_{2}=\sum_{r=1}^{4}[(\bar{\Psi}_{1})_{r}\Phi_{1}(\Psi_{1})_{r}+(\bar{\Psi}_{1})_{4+r}\Phi_{2}(\Psi_{2})_{4+r}+(\bar{\Psi}_{1})_{8+r}\Phi_{1}(\Psi_{lR})_{8+r}+(\bar{\Psi}_{1})_{12+r}\Phi_{2}(\Psi_{2})_{12+r}].
\end{eqnarray*}
However, in both cases we have to redefine the four-dimensional fields by a constant phase (which, obviously, does not modify the previous results)
\begin{equation}
\psi\rightarrow c_{\psi}\psi \qquad c^{*}_{\psi}c_{\psi}=1
\end{equation}
so that the interference between the phases of right-handed and left-handed fields determinate fermion masses 
\begin{equation}
m_{\psi}=gv_{1/2}(c^{*}_{\psi R}c_{\psi L}+c^{*}_{\psi L}c_{\psi R}).
\end{equation}
At this point, because of the arbitrariness of the coefficients $c$, the mass spectrum for all particles in Standard Model can be reproduced by setting $g$ greater than the biggest mass observed (top mass $m_{t}\simeq 180 GeV$).\\
Therefore, in this scheme masses are produced by the interference between phases of massless (right-handed and left-handed) fields and not only by the vacuum expectation value of the Higgs field.\\
We note that in the seven dimensional case, an alternative way to obtain massive fermions is simply to rewrite in our formalism mass terms in the Lagrangian density (\ref{mass}). The expression needed for this task is the following one    \begin{equation}
\Lambda_{\Psi\phi}=\sum_{l=1}^{3}\big[g_{l}(\bar{\Psi}_{lL}\Phi\Psi_{lR}+\bar{\Psi}_{lR}\Phi^{\dag}\Psi_{lL})+g_{\nu_{l}}(\bar{\Psi}_{lL}\widetilde{\Phi}\Psi_{lR}+\bar{\Psi}_{lR}\widetilde{\Phi}^{\dag}\Psi_{lL})\big].
\end{equation}
Finally, we stress again the result of this section: in our scheme it is possible to introduce a scalar field, which plays the same role of the Higgs bosons in Standard Model. The difference between them stands in the fact that in our case the scalar field has two components with opposite hypercharge; however, after the spontaneous symmetry breaking, the remaining fields Lagrangian density is exactly the same as in Standard Model.   

\section{Concluding remarks}
Therefore, we develop a model in which both gravity and electro-weak interactions are geometric; in particular, the geometrization of a $SU(2)\otimes U(1)$ gauge theory require the introduction of an extra-dimensional space and the identification of gauge with space-time isometries. In this way, gauge bosons arise as metric components.\\
We introduce spinors such that their interactions with gauge bosons are automatically contained in the Lagrangian density and their gauge charges are conserved.\\In the last section we also reproduce the spontaneous symmetry breaking mechanism.\\
Our final action is thus the sum of tree terms
\begin{equation}
S=S_{H-E}+S_{\Psi}+S_{\Phi}
\end{equation}
where \begin{itemize}\item the former is the Einstein-Hilbert action, from which the four-dimensional Einstein-Yang-Mills action outcomes; \item the second is the Dirac
action, that gives the four-dimensional
theory for spinors interacting with gauge bosons, \item the latter reproduces the
spontaneous symmetry breaking mechanism and it predicts an Higgs
field constituted by two hypercharge singlets.\end{itemize}
Even if the model is developed in a multidimensional scenario, in the low energy limit the four-dimensional chirality eigenstate arise. In fact, they are the lightest modes, since their mass does not receive any contribution due to the extra-dimensional dependence.\\ 
We have shown that there are two space-time suitable for our approach: $V^{4}\otimes S^{1}\otimes S^{2}$ and $V^{4}\otimes S^{1}\otimes S^{3}$.\\
In the first case, the geometric properties of the space-time force us to introduce 8-components spinors; thus, the necessity to deal with isospin doublets becomes natural in this scheme. This result suggest us to recast even the right-handed fields into the same geometrical object.\\  
Instead, the choice of an 8-dimensional space-time manifold lead to 16 components spinor fields; so that we propose to put into the same spinor quarks and leptons. As a consequence of this approach, the equality of the number of leptonic families and quark generations arises.\\
Furthermore, in our model we assume that extra-dimensional spinor connections vanishes. This assumption stands on the breaking of general covariance in the extra-space; in fact, a spinor does not change under translations, which are the only symmetries of $B^{k}$.\\ 
However, the breaking of general covariance itself is an open question; a possible answer could come from the Spontaneous Compactification Mechanism \cite{CS76} \cite{CS77}. A role in this sense  might also be played by the $\alpha$ fields, whose dynamics is under investigation.\\
Finally, prospectives of our work deal with the inclusion of strong interactions; as a starting point, we can say that we cannot find a space with Killing vectors able to reproduce the $SU(3)$ algebra. Furthermore, in the 8-dimensional case, the gauge group acts only some spinors components. For these reasons, the next task is the application of this approach to Grand Unification Theories (GUT).

\end{document}